\documentstyle[preprint,aps,epsf]{revtex}

\tightenlines
\draft

\title{Toy Model for Molecular Motors}

\author{Hailemariam Ambaye und Klaus W. Kehr}
\address{Institut f\"ur Festk\"orperforschung, Forschungszentrum
  J\"ulich GmbH,  D-52425 J\"ulich, Germany}
\date{\today}

\begin{document}
\maketitle

\begin{abstract}
A hopping model for molecular motors is presented consisting of a state with
asymmetric hopping rates with period 2 and a state with uniform hopping rates.
State changes lead to a stationary unidirectional current of a particle. The current is
explicitly calculated as a function  of the rate of state changes, including 
also an external bias field. 
The Einstein relation 
between the linear mobility of the particle and its diffusion coefficient is
investigated.
The power input into the system is derived, as well
as the power output resulting from the work performed against the bias field.
The efficiency of this model is found to be rather small. 

\end{abstract}
\pacs{05.40+}

\section{Introduction}

The modelling of molecular motors found great interest recently 
\cite{magn1,AB,DHR,PCPA,magn2,CAP,ZC,HB,rda}.
The aim is the understanding of the working of molecular motors, where
motor proteins perform  unidirectional motion on structures having potentials
without inversion symmetry. Several processes were proposed that may lead to 
unidirectional motion of particles in such potentials. One process is a 
switching of the potential between different states, which may be caused 
by chemical reactions \cite{PCPA,JAP}. To understand the working of
molecular motors in principle,
simplified models are desired which are amenable to explicit calculations. 
In this note we present a very simple hopping model for unidirectional 
motion, which includes switching of the hopping potential between two states. 
Due to its simplicity, all quantities of interest can be calculated explicitly.

Recently two other simplified hopping models for molecular motors were introduced.
Schimansky-Geier et al. \cite{SGKF} considered chains of segments that can 
switch either individually or collectively between two states. The segments
consist of three sites with a non-symmetric arrangement of hopping rates in 
one state and uniform hopping rates in the other state. Kolomeisky and Widom
\cite{KW} studied a model with coherent switching between two periodic 
non-symmetric potentials. In each 
potential only downward transitions are allowed; consequently the motion of a
particle comes to rest in each state in the absence of switching events.  
Our model is simpler than the model of Schimansky-Geier in that
the segments have only length 2. We restrict the derivations to coherent
switching between two states. The model is also different from the one 
of Kolomeisky and Widom in that it includes diffusion of the particle in the two states; 
unidirectional
motion is then brought about by the state changes.

We restrict in this note the derivations to the stationary state with a unidirectional
particle current; also
we do not intend to model real molecular motors with our simplified model.
In the following section the model is formulated in terms of its master equations.
In section III the stationary current is calculated. The case with an additional
uniform bias is treated in Sec.IV. The question of the validity of the Einstein relation 
between mobility and diffusion coefficient is addressed in Sec. V.
Section VI presents a derivation of the power input
and the power output when the systems performs work against the uniform bias field.
Section VII contains concluding remarks.

\section{Definition of the Model}

The model consists of a linear chain with sites to which hopping rates to the 
neighbor sites are associated. The chain exists in two different states with 
different hopping rates. 
The model is pictorially represented in Fig.1  where the hopping rates are visualized 
by potentials. In the state (1) with an asymmetric
potential, the sites with rates $\gamma_1, \delta_1$ are designated by $1$,
and the ones with $\gamma_2, \delta_2$ by $2$. The transition to the other state (2)
occurs in form of a Poisson process with rate $\Gamma_f$, and a particle
on a site of type
$1$ or $2$  is transferred to a site designated by $3$ or $4$, respectively. 
In the second state with uniform potential all sites have transition rates
$\gamma$ to the neighbor sites. The reverse transition to the non-symmetric
state occurs with the same rate $\Gamma_f$. Periodic boundary conditions will
be imposed on the system. To derive the current, it is sufficient to take
a period 2. Displacements of the particle, however, have to be counted in the extended
potentials. 

Let $P_1(t)$ be the probability of finding the particle in any of the sites
of type $1$ in the chain in state (1). The probabilities $P_2(t)$, $P_3(t)$ and
$P_4(t)$ are defined analogously. The probabilities fulfill the master equations
\begin{eqnarray}
  \label{ME}
  \frac{dP_1}{dt} &=& (\gamma_2 + \delta_2) P_2 - (\gamma_1+\delta_1) P_1
   +\Gamma_f (P_3-P_1) \nonumber \\
  \frac{dP_2}{dt} &=& (\gamma_1 + \delta_1) P_1 - (\gamma_2+\delta_2) P_2
   +\Gamma_f (P_4-P_2) \nonumber \\
  \frac{dP_3}{dt} &=& 2\gamma P_4 - 2\gamma P_3  +\Gamma_f (P_1-P_3) \nonumber \\
  \frac{dP_4}{dt} &=& 2\gamma P_3 - 2\gamma P_4  +\Gamma_f (P_2-P_4) 
\end{eqnarray}
The initial conditions are of no interest for the stationary state. 

The sum of the four equation shows that the time derivative of the sum of 
the four probabilities vanishes. Hence $P_1+P_2+P_3+P_4 = const = 1$.
The sum of the first and the second master equation, and the sum of the
third and fourth equation lead to the system of equations
\begin{eqnarray}
\label{stch}
\frac{d(P_1+P_2)}{dt} &=& \Gamma_f(P_3+P_4) -\Gamma_f(P_1+P_2) \nonumber  \\
\frac{d(P_3+P_4)}{dt} &=& \Gamma_f(P_1+P_2) -\Gamma_f(P_3+P_4) \:.
\end{eqnarray}
This set of equations shows that the state changes occur with rate
$\Gamma_f$. The stationary solution of these equations is 
$P_1+P_2 = P_3+ P_4 = \frac{1}{2}$
where we have also used the normalization of the sum of all probabilities. 

\section{Stationary Current}

The current from sites of type $1$ to $2$, or from $3$ to $4$, respectively,
in the positive direction is given by
\begin{equation}
\label{J21}
J_{21}=\gamma_1 P_1 - \delta_2 P_2 + \gamma (P_3 - P_4) \:.
\end{equation}
The current from $2$ to $1$, or from $4$ to $3$, respectively, in the
positive direction is given by
\begin{equation}
\label{J12}
J_{12}=\gamma_2 P_2 - \delta_1 P_1 + \gamma (P_4 - P_3) \:.
\end{equation}
In the stationary situation, the two currents are equal, $J_{21}=
J_{12} = J$. Equating Eqs.(\ref{J21}) and (\ref{J12}) and using the 
stationary values for the sums $P_1 + P_2$ and $P_3+P_4$, we obtain 
an equation relating the differences $P_2 - P_1$ and $P_4 - P_3$,
\begin{equation}
\label{eq1}
(\gamma_1 + \gamma_2 + \delta_1 + \delta_2)\frac{1}{2}(P_2 - P_1)
+2 \gamma (P_4 - P_3) = \frac{1}{4} (\gamma_1 - \gamma_2 + \delta_1 - \delta_2) \:.
\end{equation}
We find a second equation by subtracting the fourth from the third 
equation of the set (\ref{ME}),
\begin{equation}
\label{eq2}
\Gamma_f (P_2 - P_1) - (4\gamma + \Gamma_f) (P_4 - P_3) = 0 \:.
\end{equation}
The set of the two equations (\ref{eq1}) and (\ref{eq2}) for the two
differences is readily solved. We proceed directly to the stationary
current, which follows from the sum of Eqs. (\ref{J21}) and (\ref{J12})
by inserting the differences, and the stationary values for the sums.
The final result for the current is 
\begin{equation}
\label{curr}
J =\frac{1}{2}\: \frac{\gamma \Gamma_f (\gamma_1 + \gamma_2 - \delta_1 - 
\delta_2)} {(\gamma_1 + \gamma_2 + \delta_1 + \delta_2)(4\gamma +\Gamma_f)
+  4\gamma \Gamma_f}  \:.
\end{equation}
In the derivation of the current, there appears a term 
in the numerator proportional to $ \gamma_1 \gamma_2 - \delta_1 \delta_2 $,
see also Eq.(\ref{currb}) below. This term disappears due to the condition of detailed
balance, which is valid within each of the two states of the chain. 
Consequently, the current is proportional to $\Gamma_f$ for small $\Gamma_f$,
as it should be. We point out that the condition of detailed balance is 
fulfilled within each of the two states. The introduction of the state
change with the rate $\Gamma_f$ leads to a violation of the condition
of detailed balance when the full system is considered. This violation
is necessary for the working of the motor, cf. the discussion in Ref.\cite{JAP}.

The result shows that a unidirectional current exists for finite values of 
the flip rate $\Gamma_f$. Figure 2 presents the result for a
particular choice of the hopping
rates on the chain, as a function of  the flip rate $\Gamma_f$. The
figure also contains the results of numerical simulations of the model.
The current saturates for $\Gamma_f >> \gamma $ to the value
\begin{equation}
\label{cursat}
J_{\Gamma_f \gg \gamma} \longrightarrow
\frac{1}{2}\: \gamma \: \frac{\gamma_1 + \gamma_2 - \delta_1 - 
\delta_2} {\gamma_1 + \gamma_2 + \delta_1 + \delta_2 + 4\gamma } \:.
\end{equation}
 From the continuum models with asymmetric potentials one expects that 
the unidirectional current is maximal for an optimal flip rate and
vanishes in the limit of infinite flip rate. The saturation of 
the current at infinite flip rate is an artifact of our hopping model 
with only two sites. The model would correspond, in a continuum version,
to a model containing sections with infinite slope in the potential. 
Models with at least three sites are required to remove the artifact
of saturation. For instance, the model of Schimansky-Geier et al. \cite{SGKF}
exhibits a maximum of the current as a function of the flip rate. The vanishing 
of the current in the limit of infinite flip rate, however, requires the 
presence of infinitely many sites in a segment, i.e., the continuum limit.

\section{Uniform Bias}

It is also instructive to consider the influence of a uniform bias on the 
unidirectional motion of the particle. Then the system has to work against the
bias field and the question is which bias is sufficient to stop the motion
completely. A bias is introduced by multiplying the transition rates 
in state (1) in 
positive direction by a factor $b > 1$, $\gamma_1, \gamma_2 \rightarrow
b\gamma_1, b\gamma_2$, and the transition rates in negative direction by 
$b^{-1}$, $\delta_1, \delta_2 \rightarrow b^{-1}\delta_1, b^{-1}\delta_2$. 
In the state (2), the transition rates $\gamma $ are multiplied by $b$
for the positive direction and $b^{-1}$ for the negative direction. The
bias factor is related to an external force $F$ by $b=\exp(Fa/2k_BT)$
where $a$ is the lattice constant which is usually set unity.

The calculation of the stationary current is easily extended to the case
with bias, using the modified transition rates as given above. The result is
\begin{eqnarray}
  \label{currb}
  J &=& \frac{1}{4\Delta(b)} \Bigl( [2(b+b^{-1})\gamma + \Gamma_f]
  \lbrace 2b^2\gamma_1\gamma_2   -2b^{-2}\delta_1\delta_2 \nonumber \\
  &+&(b-b^{-1})\gamma [b(\gamma_1+\gamma_2) 
  +b^{-1}(\delta_1+\delta_2)]\rbrace \nonumber \\
  &+& (b+b^{-1})\gamma \Gamma_f
  [b(\gamma_1+\gamma_2) - b^{-1}(\delta_1+\delta_2) + 2(b-b^{-1})\gamma ] \Bigr)
\end{eqnarray}
with
\begin{equation}
  \label{D(b)}
  \Delta(b) = [b(\gamma_1+\gamma_2) + b^{-1}(\delta_1+\delta_2)]
  [2(b+b^{-1})\gamma + \Gamma_f] + 2(b+b^{-1})\gamma \Gamma_f \:.
\end{equation}
The result (\ref{curr}) is readily recovered for $b=1$.

The result for the stationary current as a function of the bias $b$ is presented 
in Fig.3, for one particular value of the flip rate $\Gamma_f$. One notices 
that the current behaves almost linearly as a function of $b$, for our choice
of the parameters. The bias value where the current vanishes can be read from
the figure. We note that this value is rather small. No analytic expression can be given because the condition $J(b) = 0$
corresponds to an equation of 6$^{th}$ order in $b$. The slope of the curve
is given by
\begin{eqnarray}
  \label{slope}
\left. \frac{dJ}{db}\right|_{b=1}&=& \frac{1}{\Delta (b=1)}[(\gamma_1\gamma_2
   + \delta_1\delta_2) (4\gamma +\Gamma_f) 
+\gamma(2\gamma+\Gamma_f)(\gamma_1+\gamma_2+\delta_1+\delta_2)
+ 2\gamma^2 \Gamma_f] \nonumber\\ &-&\frac{1}{2\Delta (b=1)^2}\: [\gamma \Gamma_f(4\gamma
  + \Gamma_f)(\gamma_1 +\gamma_2-\delta_1-\delta_2)^2 ]
\end{eqnarray}
Using the slope and the value of the current at $b=1$, an approximate prediction 
for the value of $b$ where the current vanishes can be obtained.

\section{Mobility and Diffusivity}
 
The slope of $J(b)$ at $b=1$ gives a prediction for the linear mobility of a particle 
in the stationary situation, i.e., the increase/decrease of its velocity due to the action 
of a driving force. The question arises whether the linear mobility $B$ can be related to 
a diffusion coefficient $D$ by the Einstein relation  $D= k_BTB$. 

To examine this question, we calculated the diffusion coefficient of a 
particle in the system from its asymptotic mean-square displacement 
where the squared displacement is subtracted, 
\begin{equation}
\label{msqd}
<x^2(t)>-<x(t)>^2 \longrightarrow 2\:D\:t  \qquad \gamma\:t 
\rightarrow \infty \:.
\end{equation} 
To derive the mean-square displacement, site probabilities have to be
considered that are defined on a linear chain with $2N$ sites, in the
limit $N \rightarrow \infty$. The master equations (\ref{ME}) are easily
extended to include site indices. The derivation of the mean-square 
displacement is a standard procedure, see for instance \cite{HK} or
\cite{KMW}. Because it is somewhat lengthy, it will not be reproduced 
here. 

The result for the asymptotic diffusion coefficient is 
\begin{equation}
\label{Das}
D = \frac{1}{\Delta (b=1)} [(\gamma_1\gamma_2 + \delta_1\delta_2 )
  (4\gamma +\Gamma_f)   + \gamma (2\gamma +
  \Gamma_f)(\gamma_1+\gamma_2+\delta_1+\delta_2)+ 2\gamma^2 \Gamma_f] 
\end{equation}  
The diffusion coefficient is identical to the first term of the linear 
mobility as given in Eq.(\ref{slope}). Since the second term is missing, the 
Einstein relation between mobility and diffusion coefficient does not hold
here (note that we are using units where $k_BT =1$). A violation of the 
Einstein relation is not surprising for the driven non-equilibrium, 
although stationary, system that is considered here. We point out that
the numerical difference between the mobility and the diffusion
coefficient is rather small for the parameters chosen here, it is
barely visible on the scales of Figures $3$ or $4$. 

Two limiting cases of the diffusion coefficient are of interest. First,
for small flip rates $D$ can be brought into the form 
\begin{equation}
\label{D0}
D (\Gamma_f \rightarrow 0)  = \frac{\gamma }{2}+ \frac{2}
{ \gamma_1^{-1}+\gamma_2^{-1} +\delta_1^{-1}+\delta_2^{-1} } \:.
\end{equation} 
This form of the diffusion coefficient follows from a two-state model
where $D$ is deduced from a weighted superposition of the diffusion 
coefficients in state $(2)$ (first term on the rhs of (\ref{D0})) and
of the coefficient in state $(1)$ (second term). Since the flip rate
is the same in both directions, the weights are $1/2$, and the result
that is obtained agrees with (\ref{D0}) and  is independent of the flip rate. 
Our derivations show that a two-state description is only applicable at 
small flip rates. For large flip rates, the diffusion coefficient 
saturates at
\begin{equation}
\label{Dinfty}
D(\Gamma_f \rightarrow \infty) = \gamma - \frac{2\gamma^2
  -\gamma_1\gamma_2-\delta_1\delta_2}{\gamma_1+\gamma_2+\delta_1+\delta_2 + 4 \gamma } \:.
\end{equation} 
We were not able to find a physical interpretation of this limiting 
behavior.

We also performed Monte Carlo simulations of the mean-square displacement of particles about
the stationary state. The number of particles and system size used in
the simulation are respectively  $10^4$ and $4$ with periodic 
boundary condition. Figure 4  demonstrates that $<x^2(t)>-<x(t)>^2$ is indeed proportional 
to $t$ , and the diffusion coefficient $D$ can be deduced from the
simulation results. The numerical results agree with the theoretical ones.

\section{Power Input and Output}

In this section we calculate the power input when the system is switched between two states,
with an additional bias acting on the particle, and the power output, i.e., the work performed
by the particle per unit time when it moves against the uniform bias.

Power is required to switch the system from the state with a uniform potential to the state 
with the non-symmetric potential. In our model, if a particle is on a site of type $4$ in the 
uniform potential at the time where the switching occurs, its energy will be raised when it finds 
itself on site type $2$ after the switching event. The energy follows from detailed balance in 
state $(1)$,
\begin{equation}
  \label{detb}
  \beta \epsilon = \ln \frac{\delta_2} {\gamma_1} = \ln \frac{\gamma_2}{\delta_1} \:.
\end{equation}

To obtain the power input, the average occupation of sites type $4$ at the switching event
is required. To derive this quantity, we consider the dynamics of the site occupation 
under the condition that the particle is remaining in state $(1)$, or is remaining in state
$(2)$. The respective conditional probabilities, which will be called $Q_1(t), Q_2(t)$ for
state (1) and $Q_3(t), Q_4(t)$ for state $(2)$, obey analogous master equations as 
$P_1(t), \dots , P_4(t)$, cf. (\ref{ME}), except that the terms proportional to $\Gamma_f$
are absent. The two uncoupled systems of master equations are readily solved and the 
solutions are 
\begin{eqnarray}
  \label{solu}
  Q_1(t) &=& A_1 + B_1 \exp [-(\gamma_1 +\gamma_2 +\delta_1 +\delta_2)t] \nonumber \\
  Q_2(t) &=& A_2 + B_2 \exp [-(\gamma_1 +\gamma_2 +\delta_1 +\delta_2)t] \nonumber \\
  Q_3(t) &=& A_3 + B_3 \exp (-4\gamma t) \nonumber \\
  Q_4(t) &=& A_4 + B_4 \exp (-4\gamma t) \:.
\end{eqnarray}
The constants are given as
\begin{eqnarray}
  \label{con}
  A_1 &=& \frac{\gamma_2 + \delta_2} {\gamma_1 +\gamma_2 +\delta_1 +\delta_2} \nonumber \\
  B_1 &=& \frac{(\gamma_1 + \delta_1)Q_1(0) - (\gamma_2 + \delta_2)Q_2(0)}
{\gamma_1 +\gamma_2 +\delta_1 +\delta_2} \nonumber \\
  A_2 &=& \frac{\gamma_1 + \delta_1} {\gamma_1 +\gamma_2 +\delta_1 +\delta_2} \nonumber \\
  B_2 &=& - B_1 \nonumber \\
  A_3 &=& A_4 = \frac{1}{2} \nonumber \\
  B_3 &=& -B_4 = \frac{1}{2} [Q_3(0) - Q_4(0)] \:.
\end{eqnarray}

The strategy to obtain the average occupation probabilities of the four types of sites at
the average switching times is as follows: We use as initial conditions for $Q_i(0)$ in the
solutions of the uncoupled systems of master equations the complementary occupation 
probabilities $Q_j(t_f)$ in the other states at the mean switching time $t_f$, e.g.,
$Q_1(0) =Q_3(t_f)$. The mean switching time is given by $t_f = 1/\Gamma_f$. The 
coefficients for the solutions of the uncoupled master equations are then
\begin{eqnarray}
  \label{con2}
  A_1 &=& \frac{\gamma_2 + \delta_2} {\gamma_1 +\gamma_2 +\delta_1 +\delta_2} \nonumber \\
  B_1 &=& \frac{(\gamma_1 + \delta_1)Q_3(t_f) - (\gamma_2 + \delta_2)Q_4(t_f)}
{\gamma_1 +\gamma_2 +\delta_1 +\delta_2} \nonumber \\
  A_2 &=& \frac{\gamma_1 + \delta_1} {\gamma_1 +\gamma_2 +\delta_1 +\delta_2} \nonumber \\
  B_2 &=& - B_1 \nonumber \\
  A_3 &=& A_4 = \frac{1}{2} \nonumber \\
  B_3 &=& -B_4 = \frac{1}{2} [Q_1(t_f) - Q_2(t_f)] \:.
\end{eqnarray}
We now consider the solutions Eq.(\ref{solu}) at the time $t_f$ using the coefficients
given in Eq.(\ref{con2}). They constitute sets of linear equations for the required 
coefficients $Q_j(t_f)$. The solutions of these equations are
\begin{eqnarray}
  \label{sol3}
  Q_1(t_f) &=& [ A_1 -\frac{1}{2}\exp (-\Sigma t_f) + \frac{1}{2}(A_2 -A_1 )
  \exp [ -(\gamma_1 + \gamma_2 +\delta_1 + \delta_2 )t_f]]  /DN \nonumber \\
  Q_2(t_f) &=& [ A_2 -\frac{1}{2}\exp (-\Sigma t_f) - \frac{1}{2}(A_2 -A_1 )
  \exp [ -(\gamma_1 + \gamma_2 +\delta_1 + \delta_2 )t_f]] /DN \nonumber \\
  Q_3(t_f) &=& [ \frac{1}{2} -A_1 \exp (-\Sigma t_f) - \frac{1}{2}(A_2 -A_1 )
  \exp ( -4\gamma t_f)] /DN \nonumber \\
Q_4(t_f) &=& [ \frac{1}{2}  -A_2 \exp (-\Sigma t_f) + \frac{1}{2}(A_2 -A_1 )
  \exp ( -4\gamma t_f)] /DN 
\end{eqnarray}
where
\begin{eqnarray}
  \label{con3}
  \Sigma &=& \gamma_1 + \gamma_2 +\delta_1 + \delta_2 + 4\gamma \:, \nonumber \\
  DN &=& 1 - \exp(-\Sigma t_f) \:. 
\end{eqnarray}
We have verified by numerical simulations, where the system was switched at random times
according to a Poisson process with  mean flip time $t_f$, that the expressions given above
describe the average occupations at the switching times. The above derivations were formally
made for the case of no bias. A uniform bias is immediately included in the derivations
if the transition rates $\gamma_1, \gamma_2$ are replaced by $b\gamma_1, b\gamma_2$, and
the transition rates $\delta_1, \delta_2$ by $b^{-1}\delta_1, b^{-1}\delta_2$, respectively.
The final formulae will be considered for the case including bias. 

The power input is now obtained by multiplying the occupation of  type
$4$ sites at the
flipping time with the flipping rate and the energy supplied in one flip. This energy has been given 
in Eq.(\ref{detb}) in units of $k_BT$. Hence the power input is, in units of inverse time
\begin{equation}
  \label{Pin}
  P_{in} = Q_4(t_f) \: \Gamma_f \: \epsilon/2k_BT \:.
\end{equation}

The power output is obtained from the velocity induced by the switching process by
multiplying it with the force. The force is obtained from the bias parameter as
\begin{equation}
  \label{F}
  F/k_BT = 2\ln (b)  
\end{equation}
where the expression is dimensionless due to the division with $k_BT$ and the use
of lattice constant $a=1$. Since the velocity is related to the current
by a factor $2$ in view of the normalization of $1$ particle in the
reduced system with periodic boundary conditions, we have
\begin{equation}
  \label{Pout}
  P_{out} = \pm 2J(F)\: F/k_BT \:.
\end{equation}
If the current for $F \rightarrow 0$ is negative for the choice of parameters
used (as it is the case here) we compensate the sign by multiplying
with $-1$.
The current is obtained as a function of F from Eq.(\ref{currb}) by
using Eq.(\ref{F}). 

The power input, the power output and the
efficiency of the motor, which is usually defined as the ratio of the
power output to the input, are plotted in Fig.5, 6 and 7   respectively
for three chosen flip rates as a function of the force acting on the
system. The increment of the power input is so small that it cannot be seen from the figure. For example, for $\Gamma_f=0.2$ the power
input increases from $0.199\:993\:36$ to $0.199\:994\:01$ in the range of the
force considered, from $0$ to $0.1$. The efficiency of the model that
is studied here is rather small. This is related to the fact that the
unidirectional current vanishes already for small bias fields opposing
the motion of the particles, cf. Figs.$3$ and $6$. In order to devise models
that have  a higher efficiency, more sites within a repeat unit
are required, as well as proper adjustment of the parameters.

\newpage

\section{Concluding Remarks}

In this paper we have presented a very simple model for a molecular motor
 where a particle moves unidirectionally in a hopping potential which
 is flipping between a state having a potential without inversion
 symmetry and a state with uniform potential. Such models are intended
 to mimick motion of proteins in polar filaments \cite{JAP}. The
 appeal of the model, which comprises only two different types of
 sites in the non symmetric state, is that all quantities of interest can be calculated explicitly. We 
demonstrated this by calculating the stationary current 
with and without bias which results
from the switching of the system between the two states, and by deriving
the power input and output of the system. The efficiency of the system as
a motor is rather small for the parameters that were used. It was not
the intention in any case to model real molecular motors. We also compared 
the linear mobility under the influence of a small bias field with the 
diffusion coefficient that follows from the mean-square displacement. We
observed a violation of the Einstein  relation between mobility and diffusion
coefficient. We point out that we have a stationary non-equilibrium system 
where the question of the validity of the Einstein relation is nontrivial. 
The investigation of the fluctuations in these non-equilibrium systems appears
to be of further interest.

\vskip 1cm

{\bf Acknowledgement}\\
We thank R. Sambeth for pointing out an
error of a factor 0.5 in Eq.(22) to the paper.

\newpage

\begin{center}
{\bf Figure Captions}
\end{center}

\vskip 0.5cm
\noindent Fig. 1: Pictorial representation of the model for a
molecular motor, with the
transition rates indicated. 
\vskip 0.5cm
\noindent Fig. 2: Particle current without bias as a  function of the flip rate
$\Gamma_f$ between the two states.  Line: analytical result, symbols:
simulation results. The value of the transition rates used in
the simulations  
are: $\gamma_1=\gamma_2=\gamma\:\exp(-2)$, $\delta_1=\gamma\:\exp(-4)$
, $\gamma=\delta_2=0.5$. 
Transition from state $
(1)$ to state $(2)$ and vice versa occurs with the rate
$\Gamma_f$. 

\vskip 0.5cm
\noindent Fig. 3: Particle current as  a function of the bias factor
$b$ for a fixed flip rate, $\Gamma_f=0.5$. Line: analytical
result, symbols: simulation results.

\vskip 0.5cm
\noindent Fig. 4: Mean square displacement, where  the squared mean displacement
is subtracted,  shown for different flip rates as a  function of
time. Symbols:$\ast$  simulation results for 
$\Gamma_{f}=5.0$; $\times$  simulation results  for $\Gamma_{f}=0.5$; $+$
 simulation results for  $\Gamma_{f}=0.1$; short dashes: theory for 
$\Gamma_{f}=5.0$; long dashes: theory for $\Gamma_{f}=0.5$; full line:
theory for $\Gamma_{f}=0.1$.

\vskip 0.5cm
\noindent Fig. 5: Results for the  power input for three selected
values of $\Gamma_f$. The
corresponding  values of the flip rates are indicated in the
figure. Force is measured in units of $k_B\:T$

\vskip 0.5cm
\noindent Fig. 6: Power output for three different  values of
$\Gamma_f$ in units of $k_B\:T$. The values of $\Gamma_f$ are indicated
in the figure.
\vskip 0.5cm
\noindent Fig. 7: Efficiency of the system vs force for three selected
values of $\Gamma_f$, which
 are indicated on the curves.


\begin{thebibliography}{xxx}


\bibitem{magn1}
  M.~O. Magnasco, Phys.\ Rev.\ Lett.\ {\bf 71}, 1477 (1993).
\bibitem{AB}
  R.~D. Astumian and M. Bier, Phys.\ Rev.\ Lett.\ {\bf 72}, 1766 (1994).

\bibitem{DHR}
  C.~R. Doering, W. Horsthemke, and J. Riordan, Phys.\ Rev.\ Lett.\
  {\bf 72}, 2984 (1994).
\bibitem{PCPA}
  J. Prost, J.~F. Chauwin, L. Peliti, and A. Ajdari, Phys.\ Rev.\ Lett.\
  {\bf 72}, 2652 (1994).
\bibitem{magn2}
  M.~O. Magnasco, Phys.\ Rev.\ Lett.\ {\bf 72}, 2656 (1994).
\bibitem{CAP} 
J.F. Chauwin, A. Ajdari, and J. Prost, Europhys.Lett. {\bf 27}, 421 (1994)
\bibitem{ZC}
  H.~X. Zhou and Y.~D. Chen, Phys.\ Rev.\ Lett.\ {\bf 77}, 194 (1996).
\bibitem{HB} 
P. H\"anggi and R. Bartussek, in {\sl Nonlinear Physics of Complex Systems
-- Current Status and Future Trends}, J. Parisi, S.C. M\"uller, and W. Zimmermann, Eds.,
Vol.{\bf 476}, 294 (Springer, Berlin, 1996)

\bibitem{rda} 
R.D. Astumian, Science {\bf 276}, 917 (1997)
\bibitem{JAP}
F. J\"ulicher, A. Ajdari, and J. Prost, Rev.Mod.Phys. {\bf 69}, 1269 (1997)

\bibitem{SGKF}
L. Schimansky-Geier, M. Kschischo, and T. Fricke, Phys.Rev.Lett. {\bf 79}, 3335 (1997)
\bibitem{KW}
A.B. Kolomeisky and B. Widom, J.Stat.Phys., in press

\bibitem{HK}
J.W. Haus and K.W. Kehr, Phys.Rep. {\bf 150}, 263 (1987)

\bibitem{KMW}
K.W. Kehr, K. Mussawisade, and T. Wichmann, in {\sl Diffusion in Condensed Matter},
J. Kaerger, P. Heitjans, and R. Haberlandt, Eds., Vieweg, Braunschweig, 1998, p. 265. 

\end{thebibliography}
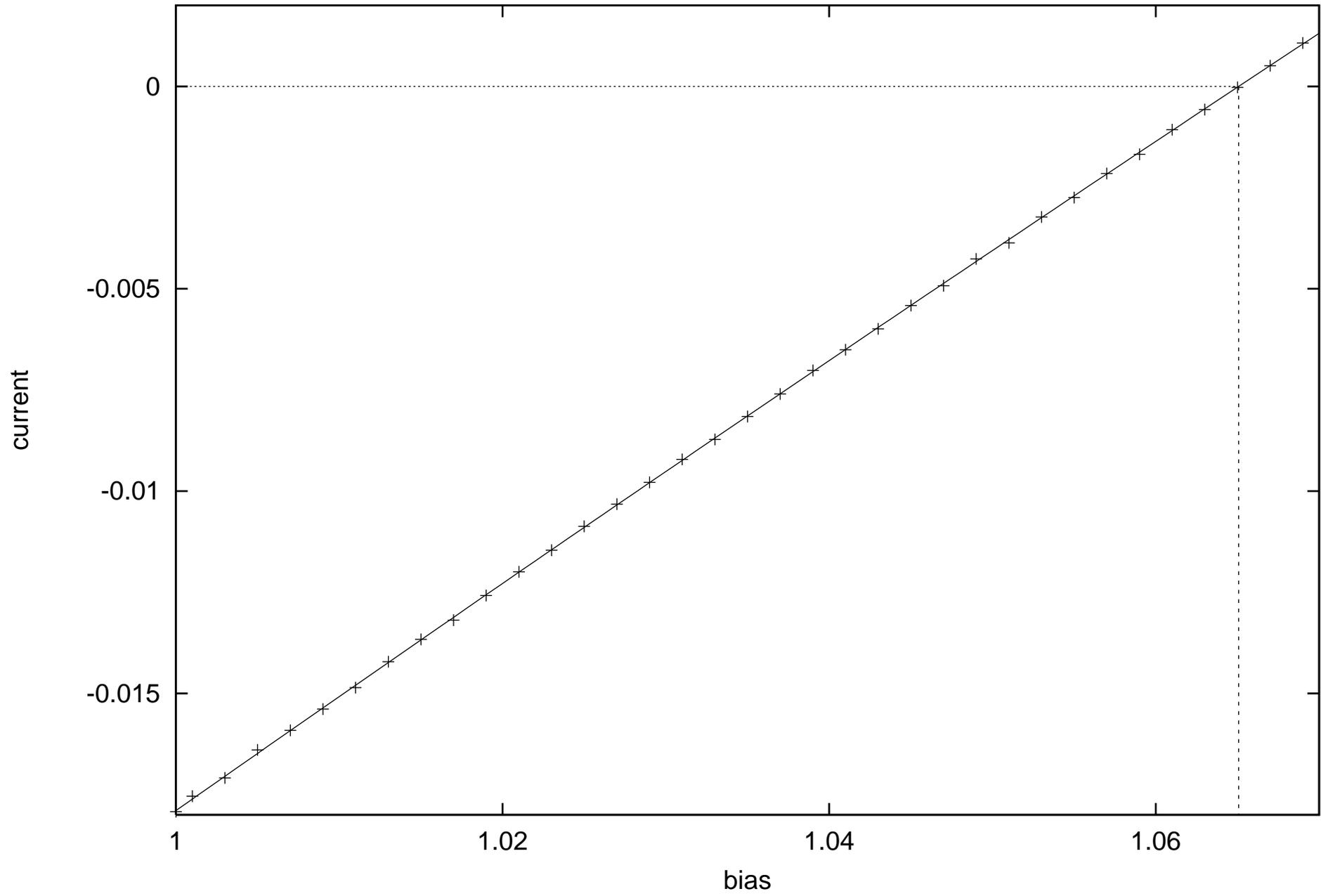
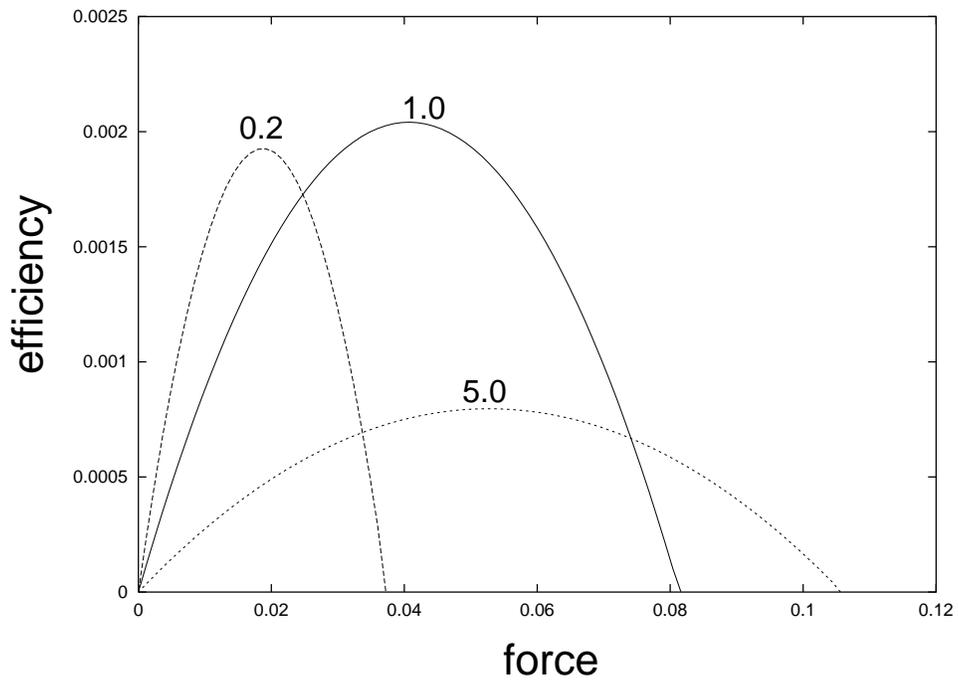
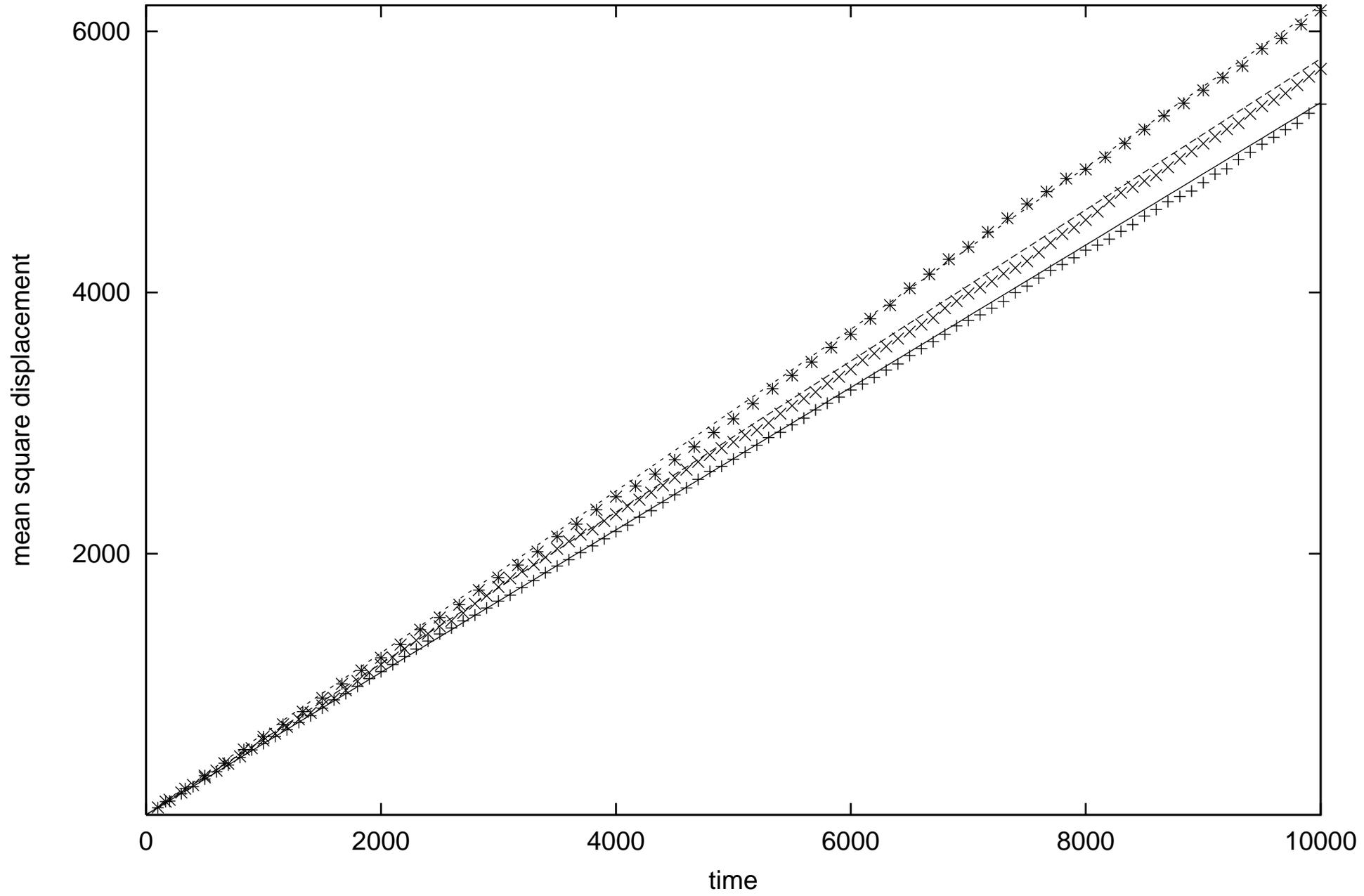
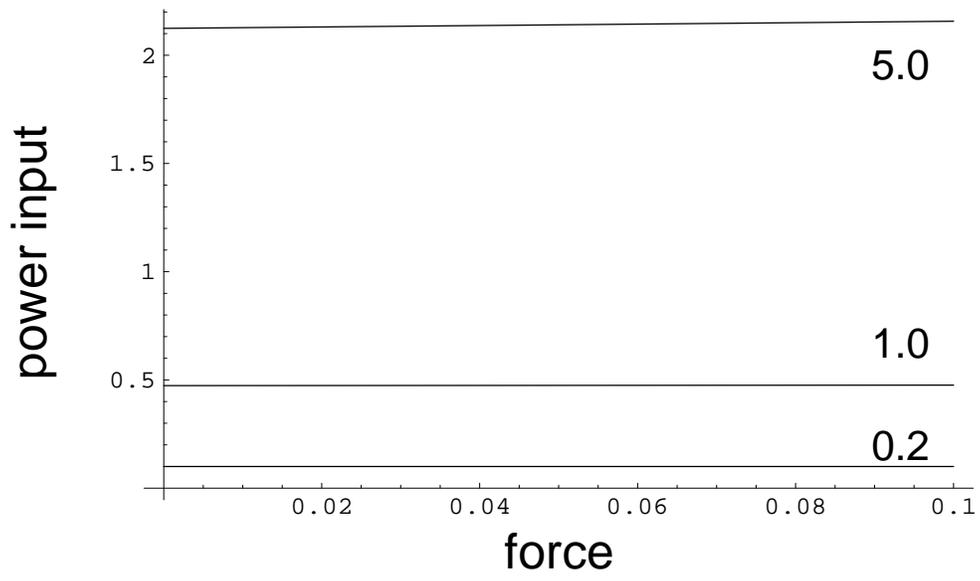
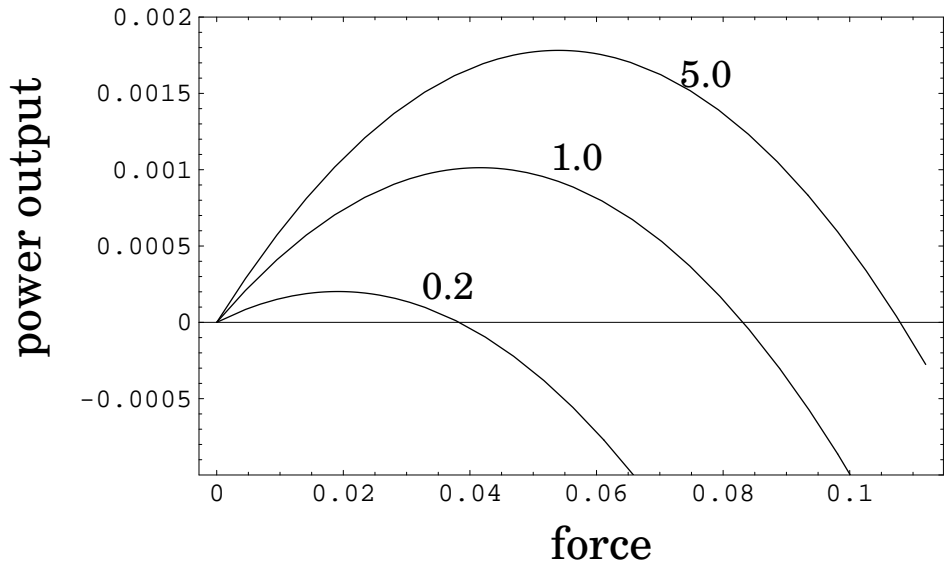
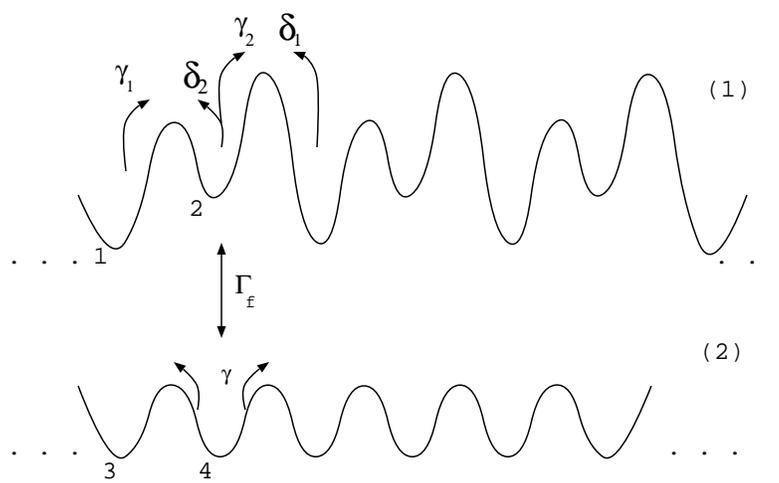
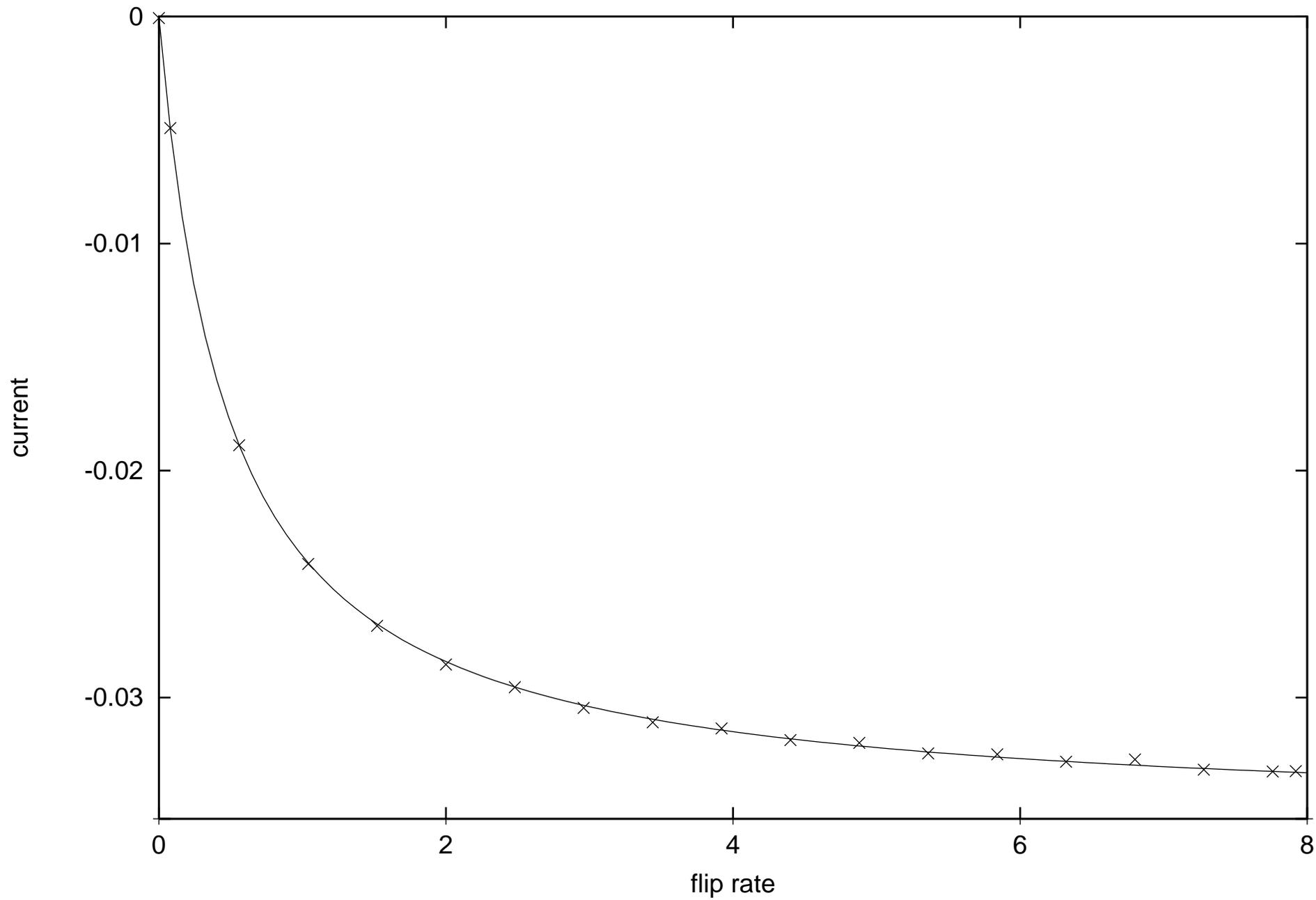
\end{document}